\documentclass[sigplan,screen]{acmart}
\acmConference[PL'20]{}{}{}
\acmYear{2020}
\acmISBN{} 
\acmDOI{} 
\startPage{1}

\setcopyright{none}

\usepackage{booktabs}   
\usepackage{subcaption} 
\usepackage{listings}
\usepackage{xargs}                      
\usepackage{pgfplots}
\pgfplotsset{width=9cm,height=6cm,compat=1.8}
\newcommand{\Csharp}{C\#}
\newcommand{\Cplusplus}{C\texttt{++}}

\newcommand{\abbrev}[1]{\textbf{#1}}
\newcommand{\mainstream}{\abbrev{mainstream}}
\newcommand{\readable}{\abbrev{readable}}
\newcommand{\idiomatic}{\abbrev{idiomatic}}
\newcommand{\documented}{\abbrev{documented}}
\newcommand{\oopatterns}{\abbrev{patterns}}
\newcommand{\common}{\abbrev{common}}
\newcommand{\expressivity}{\abbrev{expressivity}}

\begin{document}

\pagestyle{plain}

\title{GOOL: A Generic Object-Oriented Language}         
\subtitle{(extended version)}


\author{Jacques Carette}
\orcid{0000-0001-8993-9804}
\affiliation{
  \department{Department of Computing and Software}              
  \institution{McMaster University}            
  \streetaddress{1280 Main Street West}
  \city{Hamilton}
  \state{Ontario}
  \postcode{L8S 4L8}
  \country{Canada}                    
}
\email{carette@mcmaster.ca}          

\author{Brooks MacLachlan}
\affiliation{
  \department{Department of Computing and Software}              
  \institution{McMaster University}            
  \streetaddress{1280 Main Street West}
  \city{Hamilton}
  \state{Ontario}
  \postcode{L8S 4L8}
  \country{Canada}                    
}
\email{maclachb@mcmaster.ca}          

\author{W. Spencer Smith}
\orcid{0000-0002-0760-0987}
\affiliation{
  \department{Department of Computing and Software}              
  \institution{McMaster University}            
  \streetaddress{1280 Main Street West}
  \city{Hamilton}
  \state{Ontario}
  \postcode{L8S 4L8}
  \country{Canada}                    
}
\email{smiths@mcmaster.ca}          

\begin{abstract}
  We present GOOL, a Generic Object-Oriented Language.  It demonstrates
  that a language, with the right abstractions, can capture the essence of
  object-oriented programs.  We show how GOOL programs can
  be used to generate human-readable, documented and idiomatic source code in
  multiple languages.  Moreover, in GOOL, it is possible to express common
  programming idioms and patterns, from
  simple library-level functions, to simple tasks (command-line arguments, list
  processing, printing), to more complex patterns, such as methods with a
  mixture of input, output and in-out parameters, and finally Design
  Patterns (such as Observer, State and Strategy).  GOOL 
  is an embedded DSL in Haskell that can generate code in Python, Java,
  \Csharp, and \Cplusplus.
\end{abstract}



\maketitle

\section{Introduction}

Java or \Csharp? At the language level, this is close to a 
non-question: the two languages are so similar that only issues
external to the programming language itself would be the 
deciding factor.  Unlike say the question ``C or Prolog?'', which
is almost non-sensical, as the kinds of applications where each
is well-suited are vastly different.  But, given a single
paradigm, for example object-oriented (OO), would it be possible to
write a unique meta-language that captures the essence of writing
OO programs?  After all, they generally all contain (mutable)
variables, statements, conditionals, loops, methods, classes, objects,
and so on.

Of course, OO programs written in different languages appear, at
least at the surface, to be quite different. But this is mostly because
the syntax of different programming languages is different. Are they
quite so different in the utterances that one can say in them? In other
words, are OO programs akin to sentences in Romance languages
(French, Spanish, Portugese, etc) which, although different at a
surface level, are structurally very similar?

This is what we set out to explore.  One non-solution is to find an
(existing) language and try to automatically translate it to the others.
Of course, this can be made to work --- one could
engineer a multi-language compiler (such as gcc) to de-compile its Intermediate 
Representation (IR) into most of its input languages.  The end-results would 
however
be wildly unidiomatic; roughly the equivalent of a novice
in a new (spoken) language ``translating'' word-by-word.

What if, instead, there was a single meta-language designed to embody
the common semantic concepts of a number of OO languages, encoded so that the
necessary information for translation is present?  This source language could
be agnostic about what eventual target language will be used -- and free of
the idiosyncratic details of any given language.  This would be quite the boon
for the translator.  In fact, we could go even further, and attempt to
teach the translator about idiomatic patterns of each target language.

Trying to capture all the subtleties of each language is hopeless ---
akin to capturing the rhythm, puns, metaphors, similes,
and cultural allusions of a sublime poem in translation.  But programming
languages are most often used for much more prosaic tasks: writing programs
for getting things done. This is closer to translating technical textbooks,
making sure that all of the meaningful material is preserved.

Is this feasible? In some sense, this is already old hat:
modern compilers have a single IR,
used to target multiple processors. Compilers can generate
human-readable symbolic assembly code for a large family of CPUs. But this
is not the same as generating human-readable, idiomatic high-level
code.

More precisely, we are interested in capturing the conceptual meaning of
OO programs, in such a way as to fully automate the translation from
the ``conceptual'' to human-readable, idiomatic code, in mainstream
languages.

At some level, this is not new.  Domain-Specific Languages (DSL),
are high-level languages with syntax and semantics tailored to a specific
domain~\cite{mernik2005and}.  A DSL
abstracts over the details of ``code'', providing notation to
specify domain-specific knowledge in a natural manner. DSL implementations
often work via translation to a GPL for execution.  Some generate
human-readable code \cite{wang1997zephyr, mooij2013gaining, hong2012green, 
beyak2011saga}.

This is what we do, for the domain of OO programs.

We have a set of new requirements:
\begin{enumerate}
\item The generated code should be human-readable,
\item The generated code should be idiomatic,
\item The generated code should be documented,
\item The generator expresses common OO patterns.
\end{enumerate}

Here we demonstrate that all of these requirements can be met. While designing 
a generic OO language is a worthwhile endeavour, we had
a second motive: we needed a means to do exactly that as part of
our Drasil project~\cite{Drasil2019, SzymczakEtAl2016}.  The idea of Drasil is
to generate all the requirements documentation and code from expert-provided
domain knowledge.  The generated code needs to be human readable so that
experts can certify that it matches their requirements.
We largely rewrote SAGA~\cite{beyak2011saga} to create
GOOL\footnote{Available at \url{https://github.com/JacquesCarette/Drasil}
	as a sub-package.}. GOOL
is implemented as a DSL embedded in Haskell that
can currently generate code in Python, Java, \Csharp, and \Cplusplus.
Others could be added, with the implementation effort being commensurate to the
(semantic) distance to the languages already supported.

First we expand on the high-level requirements for such an endeavour, in
Section~\ref{sec:req}.  To be able to give concrete examples, we
show the syntax of GOOL in Section~\ref{sec:creating}. The details of
the implementations, namely the internal representation and the
family of pretty-printers, is in Section~\ref{sec:implementation}.
Common patterns are illustrated in Section~\ref{sec:patterns}.  We
close with a discussion of related work in Section \ref{sec:related}, plans for
future improvements in Section \ref{sec:future}, and conclusions in Section
\ref{sec:conclusions}.

Note that a short version of this paper~\cite{GOOLPEPM} will be
published at PEPM 2020. The text in both version differs many
places (other than just in length), but do not differ in meaning.

\section{Requirements} \label{sec:req}

While we outlined some of our requirements above, here we give a
complete list, as well as some reasoning behind each.

\begin{description}
\item[mainstream] Generate code in mainstream object-oriented languages.
\item[readable] The generated code should be human-readable,
\item[idiomatic] The generated code should be idiomatic,
\item[documented] The generated code should be documented,
\item[patterns] The generator should allow one to express common OO patterns.
\item[expressivity] The language should be rich enough to express a
set of existing OO programs, which act as test cases for the language.
\item[common] Language commonalities should be abstracted.
\end{description}

Targetting OO languages (\mainstream) is primarily because of their popularity,
which implies the most potential users --- in much the same way that the makers
of Scala and Kotlin chose to target the JVM to leverage the Java ecosystem, and
Typescript for Javascript.

The \readable~requirement is not as obvious. As DSL users are typically
domain experts who are not ``programmers'', why generate readable code?
Few Java programmers ever look at JVM bytecode, and few \Cplusplus{} programmers
at assembly. But GOOL's aim is different: to allow writing
high-level OO code once, but have it be available in many GPLs. One use case
would be to generate libraries of utilities for a narrow domain. As needs
evolve and language popularity changes, it is useful to have it immediately
available in a number of languages. Another use, which is core to our
own motivation as part of Drasil~\cite{SzymczakEtAl2016, Drasil2019}, is to 
have \emph{extremely well documented} code, indeed to
a level that would be unrealistic to do by hand. But this documentation is
crucial in domains where \emph{certification} is required.  And
\readable~is a proxy for \emph{understandable}, which is also quite
helpful for debugging.

The same underlying reasons for \readable~also drive \idiomatic~and \documented,
as they contribute to the human-understandability of the generated code.
\idiomatic~is important as many human readers would find the code ``foreign''
otherwise, and would not be keen on using it.
Note that documentation can span from informal comments meant for humans, to
formal, structured comments useful for generating API documentation with tools
like Doxygen, or with a variety of static analysis tools.
Readability (and thus understandability) are improved when code is 
pretty-printed~\cite{buse2009learning}. Thus taking care of layout, redundant 
parentheses,
well-chosen variable names, using a common style with lines that are not too
long, are just as valid for generated code as for human-written code.
GOOL does not prevent users from writing undocumented or complex code, if they
choose to do so. It just makes it easy to have \readable, \idiomatic~and
\documented~code in multiple languages.

The \oopatterns~requirement is typical of DSLs: common programming idioms
can be reified into a proper linguistic form instead of being merely
informal. Even some of the \emph{design patterns} of~\cite{gamma1995design}
can become part of the language itself. While this does make writing some OO
code even easier in GOOL than in GPLs, it also helps
keep GOOL language-agnostic and facilitates generating idiomatic code.
Examples will be given in Section~\ref{sec:patterns}.  But we can
indicate now how this helps: Consider Python's
ability to return multiple values with a single return statement, which
is uncommon in other languages.  Two choices might be to disallow this
feature in GOOL, or throw an error on use when generating code in languages
that do not support this feature. In the first case, this would likely mean
unidiomatic Python code, or increased complexity in the Python generator to
infer that idiom. The second option is worse still: one might have to resort
to writing language-specific GOOL, obviating the whole reason for the language!
Multiple-value return statements are always used when a function returns 
multiple
outputs; what we can do in GOOL is to support such multiple-output functions,
and then generate the idiomatic pattern of implementation in each target
language.

\expressivity~is about GOOL capturing the ideas contained in OO programs.  We
test GOOL against real-world examples from the Drasil project, such as 
software for determining whether glass withstands a nearby explosion and 
software for simulating projectile motion.

The last requirement (\common) that language commonalities be abstracted, is
internal: we noticed a lot of repeated code in our initial
backends, something that ought to be distasteful to most programmers. For
example, writing a generator for both Java and \Csharp{} makes it incredibly
clear how similar the two languages are.

\section{Creating GOOL} \label{sec:creating}

How do we go about creating a ``generic'' object-oriented language?
We chose an incremental abstraction approach: start from OO programs written in 
two different languages, and unify them \emph{conceptually}.

We abstract from concrete OO programs, not just to meet our
\expressivity~requirement, but also because that is our ``domain''.  Although
what can be said in any given OO language is quite broad, what we
\emph{actually want to say} is often much more restricted. And what we
\emph{need to say} is often even more concise.
For example, Java offers introspection features, but \Cplusplus~doesn't, so
abstracting from portable OO will not feature introspection (although it
may be the case that generating idiomatic Java may later use it);
thus GOOL as a language does not encode introspection.  \Cplusplus~templates
are different: while other languages do not necessarily have comparable
meta-programming features, as GOOL is a code generator, it is not only
feasible but in fact easy to provide template-like features, and even aspects of
partial evaluation directly. Thus we do not need to generate templates.
In other words, we are trying to abstract over the fundamental ideas
expressed via OO programs, rather than abstracting over the languages ---
and we believe the end result better captures the essence of OO programs.
Of course, some features, such as types, which don't exist per se in 
Python but are required in Java, \Csharp~and \Cplusplus, will be present
as doing full type inference is unrealistic.

Some features of OO programs are not operational: comments and formatting
decisions amongst them.  To us, programs are a bidirectional means of
communication; they must be valid, executable programs by computers, but
also need to be readable and understandable by humans.
Generating code for consumption by machines is well understood and performed by 
most DSLs, but generating code for human consumption has been given less 
attention. We tried to pay close attention to program features --- such
as the habits of programmers to write longer methods as blocks separated
by (at least) blank lines, often with comments --- which make programs
more accessible to human readers.

Finding commonalities between OO programs is most easily done from the core 
imperative language outwards.
Most languages provide similar basic types (variations on integers,
floating point numbers, characters, strings, etc.) and functions to deal
with them. The core expression language tends to be extremely similar
across languages. One then moves up to the statement language ---
assignments, conditionals, loops, etc.  Here we start to encounter
variations, and choices can be made; we'll cover that later.

For ease of experimentation, GOOL is an Embedded
Domain Specific Language (EDSL) inside Haskell.  We might eventually give GOOL 
its own external syntax, but for now it works well as a Haskell EDSL, 
especially as part of Drasil. Haskell is very well-suited
for this, offering a variety of features (GADTs, type classes,
parametric polymorphism, kind polymorphism, etc.) that are quite useful
for building languages.  Its syntax is also fairly liberal, so that
with \emph{smart constructors}, one can somewhat mimic the
usual syntax of OO languages.

\subsection{GOOL Syntax: Imperative core} \label{ssec:syntax}

Basic types in GOOL are \verb|bool| for Booleans,
\verb|int| for integers, \verb|float| for doubles, \verb|char|
for characters, \verb|string| for strings, \verb|infile| for a file
in read mode, and \verb|outfile| for a file in write mode. Lists can be
specified with \verb|listType|; \verb|listType int|
specifies a list of integers. Objects are specified using
\verb|obj| followed by a class name.

Variables are specified with \verb|var| followed by the variable name and type.
For example, \verb|var "ages" (listType int)| represents a variable called
``ages'' that is a list of integers.  For common constructions,
it is useful to offer shortcuts for
defining them; for example, the above can also be done via
\verb|listVar "ages" int|. Typical use would be
\begin{lstlisting}
let ages = listVar "ages" int in
\end{lstlisting}
so that \verb|ages| can be used directly from then on. Other means for
specifying variables is shown in Table~\ref{tab:variables}.

\begin{table}[ht]
\caption{Syntax for specifying variables}
\begin{tabular}{p{0.12\textwidth} p{0.33\textwidth}}
  \textbf{GOOL Syntax} & \textbf{Semantics} \\
  \midrule
  \verb|extVar| & for a variable from an external library \\
  \verb|classVar| & for a variable belonging to a class \\
  \verb|objVar| & for a variable belonging to an object \\
  \verb|$->| & infix operator form of \verb|objVar| \\
  \verb|self| & for referring to an object in the definition of its class \\
\end{tabular}
\label{tab:variables}
\end{table}

Note that GOOL distinguishes a variable from its value\footnote{
as befits the use-mention distinction from analytic philosophy}. To get
the value of~\verb|ages|, one must write \verb|valueOf ages|. This distinction 
is motivated by semantic considerations; it is beneficial for stricter typing 
and enables convenient syntax for \oopatterns~that translate to more idiomatic
code.

Syntax for literal values is shown in Table~\ref{tab:literals} and for
operators on values in Table~\ref{tab:operators}. Each
operator is prefixed by an additional symbol based on type. 
Boolean-valued by \verb|?|, numeric by \verb|#|, and others by \verb|$|.

\begin{table}[ht]
  \caption{Syntax for literal values}
  \begin{tabular}{p{0.12\textwidth} p{0.33\textwidth}}
    \textbf{GOOL Syntax} & \textbf{Semantics} \\
    \midrule
    \verb|litTrue| & literal Boolean true \\
    \verb|litFalse| & literal Boolean false \\
    \verb|litInt| \verb|i| & literal integer \verb|i| \\
    \verb|litFloat| \verb|f| & literal float \verb|f| \\
    \verb|litChar| \verb|c| & literal character \verb|c| \\
    \verb|litString| \verb|s| & literal string \verb|s| \\
  \end{tabular}
  \label{tab:literals}
\end{table}

\begin{table}[htb]
  \caption{Operators for making expressions}
  \begin{tabular}{p{0.12\textwidth} p{0.33\textwidth}}
    \textbf{GOOL Syntax} & \textbf{Semantics} \\
    \midrule
    \verb|?!| & Boolean negation \\
    \verb|?&&| & conjunction \\
    \verb|?||| & disjunction \\
    \verb|?<| & less than \\
    \verb|?<=| & less than or equal \\
    \verb|?>| & greater than \\
    \verb|?>=| & greater than or equal \\
    \verb|?==| & equality \\
    \verb|?!=| & inequality \\
    \verb|#~| & numeric negation \\
    \verb|#/^| & square root \\
    \verb|#|| & absolute value \\
    \verb|#+| & addition \\
    \verb|#-| & subtraction \\
    \verb|#*| & multiplication \\
    \verb|#/| & division \\
    \verb|#^| & exponentiation \\
  \end{tabular}
  \label{tab:operators}
\end{table}

Table~\ref{tab:values} shows conditional expressions and function application.
\verb|selfFuncApp| and \verb|objMethodCallNoParams| are two shortcuts for when a
method is being called on \verb|self| or when the method takes no parameters.

Variable declarations are statements, and take a variable specification
as an argument. For \verb|foo = var "foo" int|, the corresponding variable
declaration would be \verb|varDec foo|, and initialized declarations are
\verb|varDecDef foo (litInt 5)|.
Assignments are represented by \verb|assign a (litInt 5)|. Convenient
infix and postfix operators are also provided, prefixed by \verb|&|:
\verb|&=| is a synonym for \verb|assign|, and C-like
\verb|&+=|, \verb|&++|, \verb|&-=| and \verb|&~-| (the more intuitive
\verb|&--| cannot be used as \verb|--| starts a comment in Haskell).

\begin{table}[htb]
  \caption{Syntax for conditionals and function application}
  \begin{tabular}{p{0.12\textwidth} p{0.33\textwidth}}
    \textbf{GOOL Syntax} & \textbf{Semantics} \\
    \midrule
    \verb|inlineIf| & conditional expression \\
    \verb|funcApp| & function application (list of parameters) \\
    \verb|extFuncApp| & function application, for external library
    \\
    \verb|newObj| & for calling an object constructor \\
    \verb|objMethodCall| & for calling a method on an object \\
  \end{tabular}
  \label{tab:values}
\end{table}

Other simple statements include \verb|break| and \verb|continue|,
\verb|returnState| (followed by a value to return), \verb|throw| (followed by an
error message to throw), \verb|free| (followed by a variable to free from
memory), and \verb|comment| (followed by a string used as a
single-line comment).

A single OO method is frequently laid out as a sequence of blocks of
statements, where each block represents a meaningful task.  In GOOL,
\verb|block| is used for this purpose. Thus bodies are not just a
sequence of statements (as would be natural if all we cared about was
feeding a compiler), but instead a \verb|body| is a list of blocks.
A body can be used as a function body, conditional body, loop
body, etc. This additional level of organization of statements is
operationally meaningless, but represents the actual structure of OO programs
as written by humans.  This is because programmers (hopefully!) write code to
be read by other programmers, and blocks increase human-readability.
Naturally, shortcuts are provided for single-block bodies
(\verb|bodyStatements|) and for the common single-statement case,
\verb|oneLiner|.

GOOL has two forms of conditionals: if-then-else via \verb|ifCond| (
which takes a list of pairs of conditions and bodies) and
if-then via \verb|ifNoElse|.  For example:
\begin{lstlisting}
ifCond [
  (foo ?> litInt 0, oneLiner (
    printStrLn "foo is positive")),
  (foo ?< litInt 0, oneLiner (
    printStrLn "foo is negative"))]
  (oneLiner $ printStrLn "foo is zero")
\end{lstlisting}
GOOL also supports \verb|switch| statements.

There are a variety of loops: for-loops (\verb|for|), which are
parametrized by a statement to
initialize the loop variable, a condition, a statement to update the loop
variable, and a body; \verb|forRange| loops, which are given a
starting value, ending value, and step size; and \verb|forEach|
loops.  For example:
\begin{lstlisting}
for (varDecDef age (litInt 0)) (age < litInt 10) 
  (age &++) loopBody
forRange age (litInt 0) (litInt 9) (litInt 1) loopBody
forEach age ages loopBody
\end{lstlisting}
While-loops (\verb|while|) are parametrized by a condition and a body; 
try-catch (\verb|tryCatch|) is parameterized by two bodies.

\subsection{GOOL Syntax: OO features}

A \verb|function| declaration is followed by the function
name, scope, binding type (static or dynamic), type, list of parameters, and
body. Methods (\verb|method|) are defined similarly, with the addition of the
the containing class' name.  Parameters are built from
variables, using \verb|param| or \verb|pointerParam|. For example, assuming
variables ``num1'' and ``num2'' have been defined, one can define an
\textsf{add} function as:
\begin{lstlisting}
function "add" public dynamic_ int 
  [param num1, param num2]
  (oneLiner (returnState (num1 #+ num2)))
\end{lstlisting}
The  \verb|pubMethod| and \verb|privMethod| shortcuts are useful for public
dynamic and private dynamic methods, respectively. \verb|mainFunction|
defines the main function of a program. \verb|docFunc|
generates a documented function from a function description and
a list of parameter descriptions, an optional description of the return
value, and the function itself.  This generates Doxygen-style comments.

Classes are defined with \verb|buildClass| followed by the class name, name of
the class from which it inherits (if applicable), scope, list of state 
variables, and list of
methods. State variables can be built by \verb|stateVar| followed by scope, 
static or dynamic binding, and the variable itself.  \verb|constVar| can be 
used for constant state
variables. Shortcuts for state variables include \verb|privMVar| for private
dynamic, \verb|pubMVar| for public dynamic, and \verb|pubGVar| for public
static variables. For example:
\begin{lstlisting}
buildClass "FooClass" Nothing public
  [pubMVar 0 var1, privMVar 0 var2] [mth1, mth2]
\end{lstlisting}
\verb|Nothing| here indicates that this class does not have a parent,
\verb|privClass| and \verb|pubClass| are shortcuts for private and public
classes, respectively. \verb|docClass| is like \verb|docFunc|.

\subsection{GOOL syntax: modules and programs}

Akin to Java packages and other similar constructs, GOOL has modules
(\verb|buildModule|) consisting of a name, a list of libraries to import,
a list of functions, and a list of classes. Module-level comments are done
with \verb|docMod|.

At the top of the hierarchy are programs, auxiliary files, and packages. A
program (\verb|prog|) has a name and a list of files.  A \verb|package| is a
program and a list of auxiliary files; these are non-code files that
augment the program. Examples are a Doxygen configuration file
(\verb|doxConfig|), and a makefile (\verb|makefile|).  A parameter of
\verb|makefile| toggles generation of a \verb|make doc| rule, to
compile the Doxygen documentation with the generated Doxygen configuration
file.

\section{GOOL Implementation} \label{sec:implementation}

There are two ``obvious'' means of dealing with large embedded DSLs in Haskell:
either as a set of Generalized Algebraic Data Types (GADTs), or using a set of
classes, in the ``finally tagless'' style~\cite{carette2009finally} (we will
refer to it as simply \emph{tagless} from now on).  The current implementation
uses a ``sophisticated'' version of tagless. A first implementation of GOOL,
modelled on the multi-language generator SAGA~\cite{beyak2011saga} used a
straightforward version of tagless, which did not allow for enough generic
routines to be properly implemented.  This was replaced by a version based on
GADTs, which fixed that problem, but did not allow for \emph{patterns} to be
easily encoded. Thus the current version has gone back to tagless, but also uses
\emph{type families} in a crucial way.

In tagless the means of encoding a language,
through methods from a set of classes, really encodes a generalized
\emph{fold} over any \emph{representation} of the language.  Thus what
looks like GOOL ``keywords'' are either class methods or generic functions
that await the specification of a dictionary to decide on the final
interpretation of the representation.  We typically instantiate these to
language renderers, but we're also free to do various static analysis passes.

\begin{figure}[p]
\includegraphics[scale=0.289]{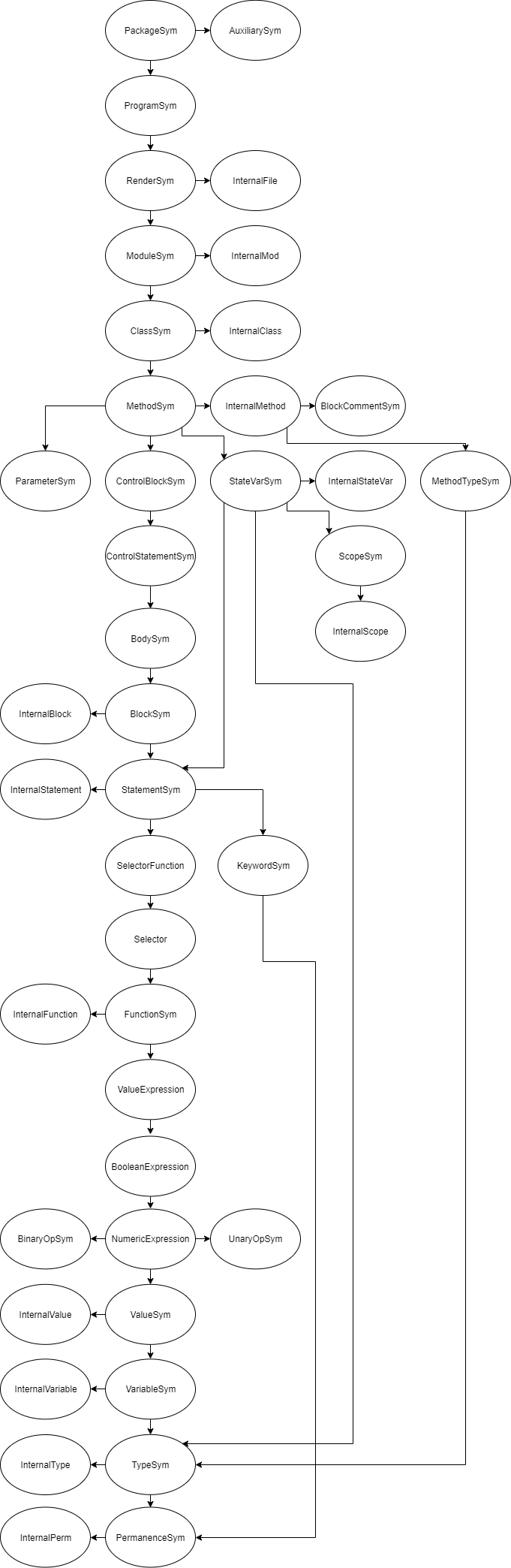}
\caption{Dependency graph of all of GOOL's type classes}
\label{fig:classes}
\end{figure}

Because tagless representations give an embedded syntax to a DSL while
being polymorphic on the eventual semantic interpretation of the terms,
\cite{carette2009finally} dubs the resulting classes ``symantic''.
Our language is defined by a hierarchy of $43$ of these symantic classes,
grouped by functionality, as illustrated in Figure~\ref{fig:classes}.  
For example, there are classes for programs,
bodies, control blocks, types, unary operators, variables, values, selectors,
statements, control statements, blocks, scopes, classes, modules, and so
on.  These define $328$ different methods --- GOOL is not a small language!

For example, here is how variables are defined:
\begin{lstlisting}
class (TypeSym repr) => VariableSym repr where
  type Variable repr
  var :: Label -> repr (Type repr) -> 
    repr (Variable repr)
\end{lstlisting}
As variables are typed, their representation must be aware of types and
thus that capability (the~\verb|TypeSym| class) is a constraint.  The
\emph{associated type}~\verb|type Variable repr| is a representation-dependent
type-level function.  Each instance of this
class is free to define its own internal representation of what a
\verb|Variable| is. \verb|var| is then a constructor for variables,
which takes a \verb|Label| and a representation of a type, returning
a representation of a variable.  Specifically, \verb|repr| has kind
\verb|* -> *|, and thus \verb|Variable| has kind \verb|(* -> *) -> *|.
In \verb|repr| \verb|(X repr)|, the type variable \verb|repr| appears
twice because there are two layers of abstraction: over the target
language, handled by the outer \verb|repr|, and over the underlying
types to which GOOL's types map, represented by the inner \verb|repr|.

We make use of this flexibility of per-target-language representation
variation to record more (or less) information for
successful idiomatic code generation. For example, the internal representation 
for a 
state variable in \Cplusplus{} stores the corresponding destructor code,
but not in the other languages.

For Java, we instantiate the \verb|VariableSym| class as follows:
\begin{lstlisting}
instance VariableSym JavaCode where
  type Variable JavaCode = VarData
  var = varD
\end{lstlisting}
where \verb|JavaCode| is essentially the \verb|Identity| monad:
\begin{lstlisting}
newtype JavaCode a = JC {unJC :: a}
\end{lstlisting}
The \verb|unJC| record field is useful for type inference: when applied to
an otherwise generic term, it lets Haskell infer that we are wishing
to only consider the \verb|JavaCode| instances.  \verb|VarData| is defined as
\begin{lstlisting}
data VarData = VarD {
  varBind :: Binding,
  varName :: String,
  varType :: TypeData,
  varDoc :: Doc}
\end{lstlisting}
Thus the representation of a (Java) variable consists of more than just its
printed representation (the \verb|Doc|~field), but also its binding time,
name, and type of the variable. \verb|Doc| comes from the package 
\verb|Text.PrettyPrint.HughesPJ| and represents formatted text. It is common in 
OO programs to declare some variables as \verb|static| to signify that the 
variable should be bound at compile-time. The \verb|Binding|, either 
\verb|Static| or \verb|Dynamic|, is thus part of a variable's representation.
That a variable is aware of its type makes the generation of declarations
simpler. The inclusion of a name, as a \verb|String|, makes generating
meta-information, such as for logging, easier.

All representing structures contain at least a \verb|Doc|. It can be considered
to be our \emph{dynamic} representation of code, from a partial-evaluation
perspective. The other fields are generally \emph{static} information used to
optimize the code generation.

We prefer generic code over representation-specific code, so there is little
code that works on \verb|VarData| directly.  Instead, there are
methods like \verb|variableDoc|, part of the \verb|VariableSym| type class,
with signature:
\begin{lstlisting}
variableDoc :: repr (Variable repr) -> Doc
\end{lstlisting}
which acts as an accessor.  For \verb|JavaCode|, it is simply:
\begin{lstlisting}
variableDoc = varDoc . unJC
\end{lstlisting}

Other uses of additional information are for uniform documentation,
builds and better arrangement of parentheses.
A common documentation style for methods is to provide a 
description of each of the method's parameters. The representation for 
\verb|Method|s stores the list of parameters, which is then used to
automate this pattern of documentation. Makefiles are often used to
compile OO programs, and this process sometimes needs to know which file
contains the main module or method.  Since GOOL includes the option of 
generating a Makefile as part of a \verb|Package|, the representation for 
a \verb|Method| and a \verb|Module| store information on whether it is 
the main method or module. Redundant parentheses are typically ignored by 
compilers, but programmers still tend to minimize them in their code ---
it makes the code more human-readable. Operator precedence is used for
this purpose, and thus we also 
store precedence information in the representations for \verb|Value|s, 
\verb|UnaryOp|erators and \verb|BinaryOp|erators to elide extra
parentheses.

Note that the \verb|JavaCode| instance of \verb|VariableSym| defines the
\verb|var| function via the \verb|varD| function:
\begin{lstlisting}
varD :: (RenderSym repr) => Label -> repr (Type repr) 
  -> repr (Variable repr)
varD n t = varFromData Dynamic n t (varDocD n)

varDocD :: Label -> Doc
varDocD = text
\end{lstlisting}
\verb|varD| is generic, i.e. works for all instances, via dispatching to other
generic functions, such as \verb|varFromData|:
\begin{lstlisting}
varFromData :: Binding -> String -> repr (Type repr) 
  -> Doc -> repr (Variable repr)
\end{lstlisting}
This method is in class \verb|InternalVariable|. Several of these
``internal'' classes exist, which are not exported from GOOL's interface.
They contain functions useful for the language renderers, but
not meant to be used to construct code representations, as they reveal too
much of the internals (and are rather tedious to use).  One important
example is the \verb|cast| method, which is never needed by user-level code,
but frequently used by higher-level functions.

\verb|varDocD| can simply be \verb|text| as \verb|Label| is an
alias for a \verb|String| -- and Java variables are just names,
as with most OO languages.

\begin{figure}[!h]
\begin{tikzpicture}[scale=0.9]
\begin{axis}[
ybar,
ymin=0,
ylabel={\# common methods},
symbolic x coords={Python/Java,Python/C\#,Python/C++ Src.,Java/C\#,Java/C++ 
Src.,C\#/C++ Src.},
xtick=data,
x tick label style={rotate=40,anchor=east},
]
\addplot coordinates {(Python/Java,163) (Python/C\#,163) (Python/C++ Src.,150) 
(Java/C\#,229) 
(Java/C++ Src., 198) (C\#/C++ Src., 202)};
\end{axis}
\end{tikzpicture}
\caption{Number of common methods between renderers}
\label{fig:common}
\end{figure}
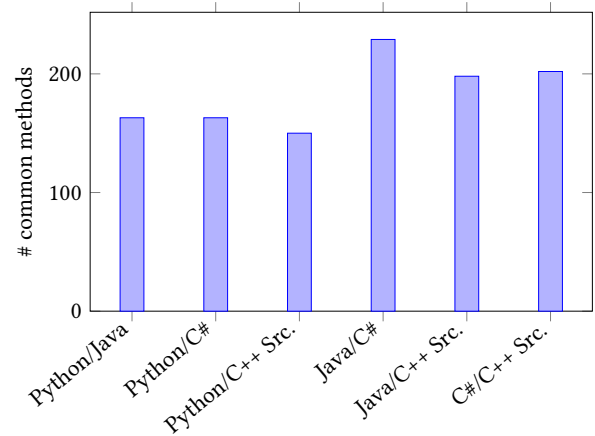

We have defined $300$ functions like \verb|varDocD|, each abstracting a 
\common ality between target languages. This
makes writing new renderers for new languages fairly straightforward.
GOOL's Java and \Csharp{} renderers demonstrate this well. Out of $328$
methods across all of GOOL's type classes, the instances of $229$ of them are
shared between the Java and \Csharp{} renderers, in that they are just calls to 
the same common function. That is $40\%$ more common instances compared to 
between Python and Java. A further $37$ instances are partially shared between 
Java and \Csharp, for example they call the same common function but with
different parameters. The numbers of common methods between each pair of 
renderers are shown in 
Figure~\ref{fig:common}. It is clear from the graph that Python is the least 
similar to the other target languages, whereas \Csharp~has the most in common 
with the others, closely followed by Java. $143$ methods are
actually the same between all $4$ languages GOOL currently targets. This might
indicate that some should be generic functions rather than class methods,
but we have not yet investigated this in detail.

Examples from Python and \Csharp{} are not shown because they both
work very similarly to the Java renderer. There are \verb|PythonCode| and
\verb|CSharpCode| analogs to \verb|JavaCode|, the underlying types are all the
same, and the methods are defined by calling common functions, where possible,
or by constructing the GOOL value directly in the instance definition, if the
definition is unique to that language.

\Cplusplus{} is different since most modules are split between a source and
header file. To generate \Cplusplus, we traverse the code twice,
once to generate the header file and a second time to generate the
source file corresponding to the same module. This is done via two instances
of the classes, for two different types: \verb|CppSrcCode| for source code and
\verb|CppHdrCode| for header code. Since a main function does not require a
header file, the \verb|CppHdrCode| instance for a module containing only a main
function is empty. The renderer optimizes empty modules/files away --- for
all renderers.

As \Cplusplus{} source and header should always be generated together, a third
type, \verb|CppCode| achieves this:
\begin{lstlisting}
data CppCode x y a = CPPC {src :: x a, hdr :: y a}
\end{lstlisting}
The type variables \verb|x| and \verb|y| are intended to be instantiated with
\verb|CppSrcCode| and \verb|CppHdrCode|, but they are left generic
so that we may use an even more generic \verb|Pair| class:
\begin{lstlisting}
class Pair (p :: (* -> *) -> (* -> *) -> (* -> *)) where
  pfst :: p x y a -> x a
  psnd :: p x y b -> y b
  pair :: x a -> y a -> p x y a

instance Pair CppCode where
  pfst (CPPC xa _) = xa
  psnd (CPPC _ yb) = yb
  pair = CPPC
\end{lstlisting}
\verb|Pair| is a \emph{type constructor} pairing, one level up from
Haskell's own \verb|(,) :: * -> * -> *|.  It is given by one constructor
and two destructors, much as the Church-encoding of pairs into the
$\lambda$-calculus.

To understand how this works, here is the instance of \verb|VariableSym|,
but for \Cplusplus:
\begin{lstlisting}
instance (Pair p) => VariableSym 
  (p CppSrcCode CppHdrCode) where
  type Variable (p CppSrcCode CppHdrCode) = VarData
  var n t = pair (var n $ pfst t) (var n $ psnd t)
\end{lstlisting}
The instance is generic in the pair representation \verb|p| but
otherwise concrete, because \verb|VarData| is concrete. The actual
instance code is straightforward, as it just dispatches to the 
underlying instances, using the generic wrapping/unwrapping
methods from \verb|Pair|.  This pattern is used for all instances,
so adapting it to any other language with two (or more) files per
module is straightforward.

At the program level, the difference between source and header is no
longer relevant, so they are joined together into a single component.
For technical reasons, currently \verb|Pair| is still used, and we arbitrarily
choose to put the results in the first component. Since generating some 
auxiliary files, especially Makefiles, requires knowledge of which are source 
files and which are header files, GOOL's representation for files stores a 
\verb|FileType|, either \verb|Source| or \verb|Header| (or \verb|Combined| for 
other languages).

GOOL's \verb|ControlBlockSym| class is worth drawing attention to. 
It contains methods for certain OO patterns, and they return \verb|Block|s, not 
\verb|Statement|s. So in addition to automating certain tasks, these methods 
also save the user from having to manually specify the result as a block.

While ``old'' features of OO languages --- basically features that
were already present in ancestor procedural languages like Algol ---
have fairly similar renderings, more recent (to OO languages) features,
such as for-each loops, show more variations.  More precisely,
the first line of a for-each loop in Python, Java, \Csharp{} and
\Cplusplus~are (respectively):
\begin{lstlisting}
for age in ages:
\end{lstlisting}
\begin{lstlisting}
for (int age : ages) {
\end{lstlisting}
\begin{lstlisting}
foreach (int age in ages) {
\end{lstlisting}
\begin{lstlisting}
for (std::vector<int>::iterator age = ages.begin(); \
  age != ages.end(); age++) {
\end{lstlisting}
where we use backslashes in generated code to indicate manually inserted
line breaks so that the code fits in this paper's narrow column margins. By 
providing \verb|forEach|, GOOL abstracts over these differences.

\section{Encoding Patterns} \label{sec:patterns}

There are various levels of ``patterns'' to encode. The previous section
documented how to encode the programming language aspects. Now we
move on to other patterns, from simple library-level functions, to
simple tasks (command-line arguments, list processing, printing), on to
more complex patterns such as methods with a mixture of input, output
and in-out parameters, and finally on to design patterns.

\subsection{Internalizing library functions}

Consider the simple trigonometric sine function, called \verb|sin| in
GOOL. It is common enough to warrant its own name, even though in most
languages it is part of a library.  A GOOL expression \verb|sin foo|
can then be seamlessly translated to 
yield \verb|math.sin(foo)| in Python, \verb|Math.sin(foo)| in Java,
\verb|Math.Sin(foo)| in \Csharp, and \verb|sin(foo)| in \Cplusplus. Other
functions are handled similarly.  This part is easily extensible, but does
require adding to GOOL classes.

\subsection{Command line arguments}

A slightly more complex task is accessing arguments passed on the command
line. This tends to differ more significantly accross languages. GOOL
offers an abstraction of these mechanisms, through an \verb|argsList| function
that represents the list of arguments, as well as convenience functions for
common tasks such as indexing into \verb|argsList| and checking if an argument
at a particular position exists. For example, these functions allow easy 
generation of code like \verb|sys.argv[1]| in Python. 

\subsection{Lists}

Variations on lists are frequently used in OO code, but the actual API
in each language tends to vary considerably; we need to provide a single
abstraction that provides sufficient functionality to do useful list
computations.  Rather than abstracting from the functionality provided
in the libraries of each language to find some common ground, we instead
reverse engineer the ``useful'' API from actual use cases.  

One thing we immediately notice from such an exercise is that lists in
OO languages are rarely \emph{linked lists} (unlike in Haskell, our host
language), but rather more like a dynamically sized vector. In particular,
indexing a list by position, which is a horrifying idea for linked lists,
is extremely common.

This narrows things down to a small set of functions and statements, shown in 
Table~\ref{tab:listfuncs}.
\begin{table}[bth]
\caption{List functions}
\begin{tabular}{p{0.15\textwidth} p{0.3\textwidth}}
  \textbf{GOOL Syntax} & \textbf{Semantics} \\
  \midrule
  \verb|listAccess| & access a list element at a given index \\
  \verb|listSet| & set a list element at a given index to a given value \\
  \verb|at| & same as \verb|listAccess| \\
  \verb|listSize| & get the size of a list \\
  \verb|listAppend| & append a value to the end of a list \\
  \verb|listIndexExists| & check whether the list has a value at a given index 
  \\
  \verb|indexOf| & get the index of a given value in a list \\
\end{tabular}
\label{tab:listfuncs}
\end{table}
For example, \verb|listAccess| \verb|(valueOf ages)| \verb|(litInt 1)| will 
generate
\verb|ages[1]| in Python and \Csharp, \verb|ages.get(1)| in Java, and
\verb|ages.at(1)| in \Cplusplus.  List slicing is a very convenient
higher-level primitive.  The \verb|listSlice| \emph{statement} gets
a variable to assign to, a list to slice, and three
values representing the starting and ending indices for the slice and the step
size. These last three values are all optional (we use Haskell's \verb|Maybe|
for this) and default to the start of the list, end of the list and $1$
respectively.  To take elements from index 1 to 2 of \verb|ages| and
assign the result to \verb|someAges|, we can use
\begin{lstlisting}
listSlice someAges (valueOf ages) (Just $ litInt 1) 
  (Just $ litInt 3) Nothing
\end{lstlisting}
List slicing is of particular note because the generated Python is particularly
simple, unlike in other languages; the Python:
\begin{lstlisting}
someAges = ages[1:3:]
\end{lstlisting}
while in Java it is
\begin{lstlisting}
ArrayList<Double> temp = new ArrayList<Double>(0);
for (int i_temp = 1; i_temp < 3; i_temp++) {
    temp.add(ages.get(i_temp));
}
someAges = temp;
\end{lstlisting}
This demonstrates GOOL's idiomatic code generation, enabled by having the
appropriate high-level information to drive the generation process.

\subsection{Printing}

Printing is another feature that generates quite
different code depending on the target language.  Here again Python
is more ``expressive'' so that printing a list (via
\verb|printLn ages|) generates \verb|print(ages)|, but in other languages
we must generate a loop; for example, in \Cplusplus:
\begin{lstlisting}
std::cout << "[";
for (int list_i1 = 0; list_i1 < \
  (int)(myName.size()) - 1; list_i1++) {
  std::cout << myName.at(list_i1);
  std::cout << ", ";
}
if ((int)(myName.size()) > 0) {
  std::cout << myName.at((int)(myName.size()) - 1);
}
std::cout << "]" << std::endl;
\end{lstlisting}
In addition to printing, there is also functionality for reading input.

\subsection{Procedures with input, output and input-output parameters}

Moving to larger-scale patterns, we noticed that our codes had methods that
used their parameters differently: some were used as inputs, some as outputs
and some for both purposes.  This was a \emph{semantic} pattern that was
not necessarily obvious in any of the implementations. However, once we noticed 
it,
we could use that information to generate better, more idiomatic code in
each language, while still capturing the higher-level semantics of the
functionality we were trying to implement.  More concretely, consider a
function \verb|applyDiscount| that takes a price and a discount, subtracts the
discount from the price, and returns both the new price and a Boolean for
whether the price is below $20$. In GOOL, using \verb|inOutFunc|, assuming
all variables mentioned have been defined:
\begin{lstlisting}
inOutFunc "applyDiscount" public static_
  [discount] [isAffordable] [price]
  (bodyStatements [
    price &-= valueOf discount,
    isAffordable &= valueOf price ?< litFloat 20.0])
\end{lstlisting}
\verb|inOutFunc| takes three lists of parameters, the input, output and
input-output, respectively.  This function has two outputs
---\verb|price| and \verb|isAffordable|--- and multiple outputs are
not directly supported in all target languages.  Thus we need to use
different features to represent these.  For example, in Python,
return statement with multiple values is used:
\begin{lstlisting}
def applyDiscount(price, discount):
    price = price - discount
    isAffordable = price < 20

    return price, isAffordable
\end{lstlisting}
In Java, the outputs are returned in an array of \verb|Object|s:
\begin{lstlisting}
public static Object[] applyDiscount(int price, \
  int discount) throws Exception {
    Boolean isAffordable;

    price = price - discount;
    isAffordable = price < 20;

    Object[] outputs = new Object[2];
    outputs[0] = price;
    outputs[1] = isAffordable;
    return outputs;
  }
}
\end{lstlisting}
In \Csharp, the outputs are passed as parameters, using the \verb|out| keyword, 
if
it is only an output, or the \verb|ref| keyword, if it is both an input and an
output:
\begin{lstlisting}
public static void applyDiscount(ref int price, \
  int discount, out Boolean isAffordable) {
    price = price - discount;
    isAffordable = price < 20;
}
\end{lstlisting}
And in \Cplusplus, the outputs are passed as pointer parameters:
\begin{lstlisting}
void applyDiscount(int &price, \
  int discount, bool &isAffordable) {
    price = price - discount;
    isAffordable = price < 20;
}
\end{lstlisting}
Here again we see how a natural task-level ``feature'', namely the
desire to have different kinds of parameters, end up being rendered differently,
but hopefully idiomatically, in each target language.  GOOL manages the
tedious aspects of generating any needed variable declarations and return
statements.  To call an \verb|inOutFunc| function, one must use
\verb|inOutCall| so that GOOL can ``line up'' all the pieces properly.

\subsection{Getters and setters}

Getters and setters are a mainstay of OO programming.  Whether these achieve
encapsulation or not, it is certainly the case that saying to an OO programmer
``variable \verb|foo| from class \verb|FooClass| should have getters and 
setters''
is enough information for them to write the code. And so it is in GOOL as well.
Saying \verb|getMethod "FooClass" foo| and \verb|setMethod "FooClass" foo|. 
The generated set methods in Python, Java, \Csharp{} and \Cplusplus{} are:
\begin{lstlisting}
def setFoo(self, foo):
    self.foo = foo
\end{lstlisting}

\begin{lstlisting}
public void setFoo(int foo) throws Exception {
    this.foo = foo;
  }
}
\end{lstlisting}

\begin{lstlisting}
public void setFoo(int foo) {
    this.foo = foo;
}
\end{lstlisting}

\begin{lstlisting}
void FooClass::setFoo(int foo) {
    this->foo = foo;
}
\end{lstlisting}
The point is that the conceptually simple ``set method'' contains a number
of idiosyncracies in each target language. These details are irrelevant for
the task at hand, and this tedium can be automated. As before, there are
specific means of calling these functions, \verb|get| and \verb|set|.

\subsection{Design Patterns}
Finally we get to the design patterns of ~\cite{gamma1995design}. GOOL
currently handles three design patterns: Observer,
State, and Strategy. 

For Strategy, we draw from partial evaluation, and ensure that the set of
strategies that will effectively be used are statically known at generation
time.  This way we can ensure to only generate code for those that will
actually be used.  \verb|runStrategy| is the user-facing function; it needs the
name of the strategy to use, a list of pairs of strategy names and bodies, and
an optional variable and value to assign to upon termination of the strategy.

For Observer, \verb|initObserverList| generates an observer for a list.  More
specifically, given a list of (initial values), it generates a declaration of
an observer list variable, initially containing the given values.
\verb|addObserver| can be used to add a value to the observer list, and
\verb|notifyObservers| will call a method on each of the observers. Currently,
the name of the observer list variable is fixed, so there can only be one
observer list in a given scope.

The State pattern is here specialized to implement \emph{Finite State Machines}
with fairly general transition functions.  Transitions happen on checking, not
on changing the state.  \verb|initState| takes a name and a state label and
generate a declaration of a variable with the given name and initial state.
\verb|changeState| changes the state of the variable to a new state.
\verb|checkState| is more complex.  It takes the name of the state variable, a
list of value-body pairs, and a fallback body; and it generates a conditional
(usually a switch statement) that checks the state and runs the corresponding
body, or the fallback body, if none of the states match.

Of course the design patterns could already have been coded in GOOL, but
having these as language features is useful for two reasons: 1) the GOOL-level
code is clearer in its intent (and more concise), and 2) the resulting code
can be more idiomatic.

Below is a complete example of a GOOL function.  The recommended style is
to name all strings (to avoid hard-to-debug typos) and variables, then
write the code proper.
\begin{lstlisting}
patternTest :: (MethodSym repr) => repr (Method repr)
patternTest = let 
 fsmName = "myFSM"
 offState = "Off"
 onState = "On"
 noState = "Neither"
 obsName = "Observer"
 obs1Name = "obs1"
 obs2Name = "obs2"
 printNum = "printNum"
 nName = "n"
 obsType = obj obsName
 n = var n int
 obs1 = var obs1Name obsType
 obs2 = var obs2Name obsType
 newObs = extNewObj obsName obsType []
    
 in mainFunction (body [block [
  varDec n,

  initState fsmName offState, 
  changeState fsmName onState,
  checkState fsmName 
  [(litString offState, oneLiner $ printStrLn offState), 
   (litString onState, oneLiner $ printStrLn onState)] 
  (oneLiner $ printStrLn noState)],

  block [
   varDecDef obs1 newObs, 
   varDecDef obs2 newObs],

  block [
   initObserverList obsType [valueOf obs1], 
   addObserver $ valueOf obs2,
   notifyObservers (func printNum void []) obsType]])
\end{lstlisting}

\section{Related Work} \label{sec:related}

We divide the Related Work into the following categories
\begin{itemize}
\item General-purpose code generation
\item Multi-language OO code generation
\item Design pattern modeling and code generation
\end{itemize}
which we present in turn.

\subsection{General-purpose code generation}

\textbf{Haxe}~\cite{Haxe} is a general-purpose multi-paradigm language and 
cross-platform
compiler.  It compiles to all of the languages GOOL does, and many
others.  However, it is designed as a more traditional programming language, and
thus does not offer the high-level abstractions that GOOL provides. Furthermore
Haxe strips comments and generates source code around a custom framework; 
the effort of learning this framework and the lack of comments makes the 
generated
code not particularly readable. The internal organization of Haxe does not seem
to be well documented.

\textbf{Protokit}~\cite{kovesdan2017multi} is a DSL and code generator for Java 
and
\Cplusplus, where the generator is designed to produce
general-purpose imperative or object-oriented code. The Protokit generator is
model-driven and uses a final ``output model'' from which actual code can be
generated. Since the ``output model'' is quite similar to the generated
code, it presented challenges with regards to semantic, conventional, and
library-related differences between the target languages
\cite{kovesdan2017multi}. GOOL's finally-tagless approach and syntax for
high-level tasks, on the other hand, help overcome differences between
target languages.

\textbf{ThingML}~\cite{harrand2016thingml} is a DSL for model-driven engineering
targeting C, \Cplusplus, Java, and JavaScript. It is specialized to deal with
distributed reactive systems (a nevertheless broad range of application 
domains).
This means that this is not quite a general-purpose DSL, unlike GOOL.
ThingML's modelling-related syntax and abstractions stand in contrast to GOOL's
object-oriented syntax and abstractions. The generated code lacks some of the
pretty-printing provided by GOOL, specifically indentation, which detracts from
readability.

\subsection{Object-oriented generators}

There are a number of code generators with multiple target OO languages,
though all are for more restricted domains than GOOL, and thus do not meet all
of our requirements.

\textbf{Google protocol buffers}~\cite{Protobuf} is a DSL for serializing
structured data, which can be compiled into Java, Python, Objective C, and
\Cplusplus.  \textbf{Thrift}~\cite{slee2007thrift} is a Facebook-developed tool
for generating code in multiple languages and even multiple paradigms based on
language-neutral descriptions of data types and interfaces.
\textbf{Clearwater}~\cite{swint2005clearwater} is an approach for implementing
DSLs with multiple target languages for components of distributed systems.  The
\textbf{Time Weaver} tool~\cite{de2004glue} uses a multi-language code
generator to generate ``glue'' code for real-time embedded systems.  The domain
of mobile applications is host to a bevy of DSLs with multiple target
languages, of which \textbf{MobDSL}~\cite{kramer2010mobdsl} and
\textbf{XIS-Mobile}~\cite{ribeiro2014xis} are two examples.
\textbf{Conjure}~\cite{Conjure} is a DSL for generating APIs. It reads YML
descriptions of APIs and can generate code in Java, TypeScript, Python, and
Rust.

\subsection{Design Patterns}

A number of languages for modeling design patterns have been developed. The
\textbf{Design Pattern Modeling Language} (DPML)~\cite{mapelsden2002design} is 
similar
to the Unified Modeling Language (UML) but designed specifically to overcome
UML's shortcomings so as to be able to model all design patterns. DPML consists 
of
both specification diagrams and instance diagrams for instantiations of design
patterns, but does not attempt to generate actual source code from the models.
The \textbf{Role-Based Metamodeling Language}~\cite{kim2003uml} is also based 
on UML but
with changes to allow for better models of design patterns, with specifications
for the structure, interactions, and state-based behaviour in patterns. Again,
source code generation is not attempted. Another metamodel for design patterns
includes generation of Java code \cite{albin2001meta}. IBM developed a DSL in 
the form of a visual user interface for generation of OO code based on design 
patterns \cite{budinsky1996automatic}.  The languages that
generate code do so only for design patterns, not for any general-purpose code,
as GOOL does.

\section{Future Work} \label{sec:future}

Currently GOOL code is typed based on what it represents:
variable, value, type, or method, for example. The type system does not
go ``deeper'', so that variables are untyped, and values (such as booleans
and strings) are simply ``values''.  This is sufficient to allow us to
generate well-formed code, but not to ensure that it is well-typed.
For example, it is unfortunately possible to pass a value that is known
to be a non-list to a function (like \verb|listSize|) which requires it.
This will generate a compile-time error in generated Java, but a run-time error
in generated Python.  We have started to statically type GOOL, by making
the underlying representations for 
GOOL's \verb|Variable|s and \verb|Value|s Generalized Algebraic Data Types
(GADTs), such as this one for \verb|Variable|s:
\begin{lstlisting}
data TypedVar a where
  BVr :: VarData -> TypedVar Boolean
  IVr :: VarData -> TypedVar Integer
  ...
\end{lstlisting}
This will allow variables to have different types, and Haskell will catch
these. We would be re-using Haskell's type system to catch (some) of the
type errors in GOOL.  Because we do not need to type arbitrary code in any
of the target languages, but only what is expressible in GOOL, we can
engineer things so as to encode quite a wide set of typing rules.

GOOL is currently less-than-precise in the list of generated import statements;
we want to improve the code to track precise dependencies, and only generate
imports for the features we actually use. This could be done via weaving
some state at generation-time for example.  In general, we can do various 
kinds of static analyses to help enhance the code generation quality.
For example, we ought to be much more precise about \verb|throws Exception|
in Java.

Another important future feature is being able to interface to external 
libraries, instead of just already-known libraries. In particular, we have a 
need to
call external Ordinary Differential Equation (ODE) solvers, since Drasil 
currently focuses on scientific applications. We do not
want to restrict ourselves to a single function, but have a host of
different functions implementing different ODE-solving algorithms available.
The structure of code that calls ODE solvers varies considerably, so that we 
cannot
implement this feature with current GOOL features.  In general, we believe
that this requires a multi-pass architecture: an initial pass to collect
information, and a second to actually generate the code.

Some implementation decisions, such as the use of \verb|ArrayList| to represent
lists in Java, are hard-coded. But we could have used \verb|Vector| instead.
We would like such a choice to be user-controlled. Another such decision
point is to allow users to choose which specific external library to use.

And, of course, we ought to implement more of the common OO patterns.

\section{Conclusion} \label{sec:conclusions}

We currently successfully use GOOL to simultaneously generate code in all of 
our target languages for the glass and projectile programs described in Section 
\ref{sec:req}. 

Conceptually, mainstream object-oriented languages are similar enough that it
is indeed feasible to create a single ``generic'' object-oriented language that
can be ``compiled'' to them.  Of course, these languages are syntactically
quite different in places, and each contains some unique ideas as well.
In other words, there exists a ``conceptual'' object-oriented language that
is more than just ``pseudocode'': it is a full-fledged executable language
(through generation) that captures the common essence of mainstream OO
languages.

GOOL is an unusual DSL, as its ``domain'' is actually that of object-oriented
languages. Or, to be more precise, of conceptual programs that can be
easily written in languages containing a procedural code with an
object-oriented layer on top --- which is what Java, Python, \Cplusplus{} and
\Csharp{} are.

Since we are capturing \emph{conceptual programs}, we can achieve
several things that we believe are \emph{together} new:
\begin{itemize}
\item generation of idiomatic code for each target language,
\item turning coding patterns into language idioms,
\item generation of human-readable, well-documented code.
\end{itemize}

We must also re-emphasize this last point: that for GOOL, the generated code
is meant for human consumption as well as for computer consumption. This is
why semantically meaningless concepts such as ``blocks'' exist: to be able
to chunk code into pieces meaningful for the human reader, and provide
documentation at that level as well.


\bibliography{References}

%
%
\end{document}